# Uncertain for A Century:

## Quantum Mechanics and The Dilemma of Interpretation

Adam Frank
University of Rochester
Department of Physics and Astronomy
afrank@pas.rochester.edu

**Introduction:** Quantum Mechanics, the physical theory describing the microworld, represents one of science's greatest triumphs.  It lies at the root of all modern digital technologies and offers unparalleled correspondence between prediction and experiments.  Remarkably, however, after more than 100 years it is still unclear what quantum mechanics *means* in terms of basic philosophical questions about the nature of reality.  While there are many *interpretations* of the mathematical machinery of quantum physics, there remains no experimental means to distinguish between most of them.

In this contribution, (based on a discussion at the NYAS), I wish to consider the ways in which the enduring lack of an agreed upon interpretation of quantum physics influences a number of critical philosophical debates about physics and reality.  I briefly review two problems effected by quantum interpretations: the meaning of the term "Universe" and the nature of consciousness.  In what follows I am *explicitly not advocating for any particular quantum interpretation*.  Instead, I am interested in how the explicit inability of modern physics to experimentally distinguish between interpretations with wildly divergent ontological/epistemological implications plays into discussions of physics and its description of the world.

**Quantum Mechanics and "Bits of Matter":** Most physics students come to their college studies expecting the discipline to provide an exacting account of nature such as they were exposed to in their high school Newtonian mechanics based classes.  Most are, therefore, shocked when they're introduced to quantum mechanics and its description of the world.  In place of a clear vision of "little bits of matter", quantum physics gives us a beautiful, powerful and yet seemly paradoxical calculus.  With its emphasis on probability waves, essential uncertainties and measurements fundamentally disturbing the reality they seek to measure, quantum mechanics makes imagining matter in a conventional sense all but impossible.

This situation arose at beginning of the last century when physicists found Newtonian mechanics failed in describing the micro-world of molecules, atoms and their constituents.  In a burst of creativity, they devised a new set of rules embodied in what was called *Schrodinger's Equation*.  Like Newton's *F=ma*, Schrodinger's Equation represented the mathematical machinery for describing how matter responds to forces but with an important twist.  What drops out of Schrodinger's equation is not the

Newtonian state of exact position and velocity but something called the wave-function (physicists refer to it as $\psi$ or "psi"). Unlike the Newtonian state that can be clearly imagined in a common sense way, the wave function turned out to pose serious epistemological and ontological challenges.

While it was easy to imagine the "thing" described in Newtonian physics (a particle at a specific location with a specific velocity), the wave function leads to probabilities living at the root level of reality. In that sense $\psi$ appears to tell you that any moment in time, the particle has many positions and many velocities. In effect the "little bit of matter" so essential to Newtonian accounts of particles was smeared out into set of potentials or possibilities whose total evolution is controlled by the Schrodinger equation.

But it's not just position and velocity that get smeared out. The wave-function treats all properties of the particle in the same way – electric charge, energy, the direction the particle's spin. They all become probabilities *holding many possible values at the same time*. Taken at face value, it's as if the particle doesn't have definite properties at all. This is what Werner Heisenberg, one of the founders of quantum mechanics, meant when he advised people not to think of atoms as "things".

But this is not the end of the trip down the rabbit hole. According to the standard way of treating the calculus of quantum mechanics, *the act of making a measurement* on the particle kills off all pieces of the wave function except the one instruments actually register. The wave-function is said to *collapse* as all the smeared out, potential positions or velocities vanish in the act of measurement. It's as the Schrodinger equation's descriptive authority is abruptly halted because a measurement occurs.

Thus physicists (and philosophers) were left with two kinds of strangeness to deal with in their effort to understand what quantum mechanics tells us about the world. What exactly does the wave function describe? What really occurs in a measurement? Remarkably, even after 100 years, a path through these difficulties has not been found.

**Quantum Mechanics and Its Interpretations:** There are many interpretations of quantum theory (Leifer 2014). The earliest one to gain force, the Copenhagen Interpretation, is associated with Danish physicist Neil's Bohr and other founders of the discipline. It was meaningless, in their view, to speak of the properties of atoms in-and-of-themselves. Quantum Mechanics was a theory that spoke only to our knowledge of the world. The measurement problem highlighted this barrier between epistemology and ontology by making our role in gaining knowledge explicit. Others were not so willing to give up on an objective access to an objective world. While some hoped that so-called *hidden variables* would be discovered that made sense of quantum weirdness, others took a different view. In the so-called *Many Worlds Interpretation* (Everett 1957, 1973) the authority of the wave function and its governing Schrodinger's Equation was taken as absolute. Measurements don't suspend the equation or collapse the wave-function, it merely makes the Universe split off into many parallel versions of itself.

The problem with all these interpretations is that remains no way to experimentally distinguish between them. Which one you choose becomes a matter of philosophical temperament. On one side there are the *psi-ontologists* who want the wave function to describe the objective world "out there". On the other side there are the *psi-epistemologists* who see the wave-function as a description of our knowledge (and its limits: Fuchs & Schack 2011). The battle-lines between these positions can be fierce precisely because there is no way to answer this scientific question with science (at least for now).

It's worth noting that theoretical "no-go" theorems can be quite helpful in positing limits that different interpretations would impose on experiments. A recent example is the work of Pusey, Barrett & Rudolph (PBR 2011) who showed that certain classes of *psi-epistemological* interpretations can be ruled out (Harrigan & Spekkens 2010). The PBR theorem showed that one can't posit the existence of underlying ontic states (meaning real properties of the world) while also positing that the wave function doesn't fully represent those states (meaning they were epistemic). It is noteworthy, however, that the PBR theorem does not address what might be called "fundamentally epistemic" interpretations such as Copenhagen or neo-Copenhagen models (i.e. QBism see below)

**Quantum Interpretations and its Discontents I. The Meaning of "Universe".**

As discussed, the Many-Worlds interpretation of quantum mechanics finds favor with physicists and philosophers who advocate for an ontic view of their equations. For them all fundamental equations of mathematical physics capture aspects of a timeless reality belonging to the Universe itself. This position is deeply ingrained in the history of physics. Indeed the philosophical basis of the positions extends as far back as the Platonic Doctrine of Ideals whereby the world we perceive is but a shadow of incorruptible mathematical forms constituting the true nature of reality.

Given the enormous success of mathematical physics in revealing unseen aspects of the world's behavior, it is easy to understand the emotional force of this kind of *wave-function realism* (Ney & Albert 2013*)*. As one example, consider how Maxwell's equations for electromagnetics revealed the presence of new and previously unimagined forms of "light" such as radio waves.

But dealing with the strangeness of quantum mechanics means that each interpretation comes with its own price. Each one forces adherents to take a long step backwards from the kind of "naive realism" possible with the Newtonian word-view. Thus the Many Worlds Interpretation's ability to keep reality in the mathematical physics – the wave equation – means it's adherents must accept infinite numbers of parallel universes that are infinitely splitting off into an infinity of other parallel universes. While the idea can seem rich and exiting to some, it raises a fundamental question about the very subject considered by Physics. Science is taken to be the study of the reality we have access to via our senses and the instruments we build to extend the reach of those senses.

Without experimental justification, adherents of the Many Worlds' interpretation are willing to enlarge the definition of Universe to include realities not directly accessible to our instruments. This is done to satisfy a pre-existing attitude about reification of the "mathematical objects" comprising physical theory.

I raise this point not to argue that the Many Worlds interpretation is wrong. Instead the point is that, in the absence of an experimentally verified interpretation, it appears we remain uncertain about something as fundamental as the essential subject matter of physics. Do our equations describe the Universe we see or do they describe a Universe of infinitely larger and, perhaps, unobservable possibilities that must be considered real? Note this problem is fundamentally distinct from discussions about a Universe of infinite spatial extent of which our observable domain constitutes a sub-region.

**Quantum Interpretations and its Discontents II. What Matter For Mind?**

A second point where quantum interpretations become relevant is the relationship between mind and matter (Chalmers 2002). In the fault lines for this debate *materialism* holds a particular kind of high ground in, at least, the public versions of discussions over that the nature of consciousness. What is particularly interesting is that when taking on the problem of Mind and Brain, advocates for a cosmos fully reducible to matter often take a position embodying a kind "hard-nosed" realism. It is in light of such confidence that those who advocate the alternative - that Mind might be something more than "nothing but bits of matter" – can be cast as victims of wishful thinking, imprecise reason, or worse, an adherence to the domains of mystical "woo".

But the unanswered question of quantum interpretations leaves materialists in an unexpectedly shaky position. Rather than enjoying the high ground of metaphysical certainty, they are left with hard choices about which kinds of strangeness they're willing to swallow as we have already encountered with the Many Worlds interpretation.

Of course there is a big price to pay for the fundamental psi-epistemologist positions too. Physics from their perspective is not longer a description of the world in-and-of itself. Instead it's a description of the rules for our interaction with the world. A particularly cogent new version of the psi-epistemological position, called Quantum Bayseanism or QBism (Fuchs, Mermin & Schack 2014), raises this perspective to new levels of specificity by taking the probabilities in quantum mechanics at face value. According to QBism, the irreducible probabilities in quantum mechanics tell us it's really a theory about making bets on the world's behavior (via our measurements) and then updating our knowledge after those measurements are done. In this way QBism points explicitly to our failure to include the *observing subject* that lies at root of quantum weirdness. As physicist David Mermin recently wrote in *Nature* "QBism attributes the muddle at the foundations of quantum mechanics to our unacknowledged removal of the perceiving subject from physical science." (Mermin 2014)

How does the dichotomy between psi-epistemological and psi-ontic positions play into debates about consciousness? The answer can be found in the definition of the Hard Problem of Consciousness first articulated by David Chalmers (Chalmers 2002). Following work by Thomas Nagel, Chalmers pointed to the vividness – the intrinsic presence - of the perceiving subject's experience as a problem that no current explanatory account of consciousness seems capable of embracing.

While some see the Hard Problem as real but inherently unsolvable, others posit a range of options for its account. Consciousness may, for example, be an example of the emergence of a new entity in the Universe not contained in the laws of particles. There is also the more radical possibility that some rudimentary form of consciousness must be added to the list of things the world is build of, like mass or electric charge. But for Daniel Dennett (2002) and other who hold a strict materialist view these ideas are dead ends. As Michu Kaku wrote in his recent book "…there is no such thing as a Hard problem" (Kaku 2014). For this kind of public materialism, consciousness is just programing that sits on top of wiring that sits on top of "bits of matter".

It is however hard to reconcile this kind of easy materialist dismissal of the Hard Problem with quantum mechanics and its multiple interpretations. The reasons why Dennett and Kaku's kind of position becomes shaky are twofold and have nothing to with whether or not Newtonian mechanics might be fine for explaining the activity of the brain.

First, the difference between the psi-ontological and psi-epistemological positions is so fundamental that without knowing which one is correct it's impossible to know what quantum mechanics is intrinsically referring to. Imagine for a moment that something like the QBist interpretation of quantum mechanics were true. If this emphasis on the observing subject was the correct lesson to learn from our most powerful description of matter, then the standard view of materialism and its perspective on Mind and Matter would seem to lose its force. If QBism were true, there may be enormous surprises waiting for us in our exploration of subject and object, including additions to what entities must be included in any account of Mind. Dennett's kind of materialism - being a particular form of psi-ontology – would by necessity be blind to these kinds of additions.

Indeed, without an experimentally verifiable interpretation of quantum physics, the role of "perspective" in physics remains an interesting open issue. This comes because measurement plays such a central role *in what is not understood* about quantum theory. This is quite different from a classical theory like Relativity. There observers also take a central position but in an objectively deterministic framework.

Indeed philosopher of physics Michael Bitbol (REF) uses the locus of uncertainty surrounding quantum interpretations as one line of argument that conscious experience must treated as primary. By "primary" he means all codifications scientific knowledge begin via abstractions of experience.  As he puts it, "Conscious experience *as a whole* can by no means be reduced to structure" where structure implies sets of laws and their initial conditions.  Bitbol's presents six arguments that conscious experience cannot derive from a material basis, the last of which rests on the philosophy of quantum mechanics.

**Conclusion:** Given the difficulties discussed in the paper, one has to ask why some of the weird alternatives suggested by quantum interpretations are to be preferred over weird alternatives suggested by others.  Why does the infinity of (potentially unobservable) parallel Universes in the Many Worlds Interpretation get associated with the sober, hard-nosed position a 'la Dennett (who favors it), while including the perceiving subject is condemned being anti-scientific or an embrace of mysticism?

Indeed, in the absence of experimental evidence we are left with a range of possible visions for quantum mechanics' implications for reality that a tally of votes at the next American Physical Society, (or American Philosophical Society), won't resolve. While some interpretations may be more complete and coherent than others (this is the good work of good philosophy) in the end we remain bereft of our quaint imaginings about little bits of matter.

In this article I have pointed out two examples of how the lack of an accepted or, better yet, experimentally verified interpretation of quantum mechanics leads to an extreme range of philosophically divergent positions concerning the nature of reality and our place in it.   One of the many lessons to be learned from this situation is the caution that must be taken concerning strident metaphysical positions and claims they are based on "what physics tells us".  In truth, the dogged agnosticism of quantum mechanics to its own philosophical implications leave everyone – materialists, physicalists and idealists – on remarkably shaky ground.

Bitbol, M., 2014, "Is Consciousness Primary", http://philsci-archive.pitt.edu/4007/

Chalmers, D, 2002, "Philosphy of Mind", Oxford University Press, New York, New York.

Chalmers, D, 2002, "Consciousness and its Place in Nature", in Philosophy of Mind, Oxford University Press, New York, New York, 247

Everett, H. 1957, Rev. Mod. Phys, 29, 454

Everett, N. 1973, in The Many-Worlds Interpretation of Quantum Mechanics, ed. DeWitt, B. S. & Graham, N. (Princeton: Princeton Univ. Press)

Dennett, D, 2002, "Quining Qualia", in Philosphy of Mind, Oxford University Press, New York, New York, 226

Fuchs CA, Mermin ND, Schack R. An introduction to QBism with an application to the locality of quantum mechanics. American Journal of Physics 2014; 82(8):749–754.

Fuchs, C. A.; Schack, R. (2011). "A Quantum-Bayesian route to quantum-state space". *Foundations of Physics* **41** (3): 345–56

Harrigan, N & Spekkens, R., (2010). "Einstein, incompleteness, and the epistemic view of quantum states". *Foundations of Physics* **40**: 125

Kaku, M., 2014, The Future of the Mind: The Scientific Quest to Understand, Enhance, and Empower the Mind, Doubleday, New York

Leifer, M, (2014), "Is the Quantum State Real?", Quanta, 3, 67

Mermin ND. (2013), "Annotated interview with a QBist in the making", In Elegance and Enigma: The Quantum Interviews, Schlosshauer M (editor). Berlin: Springer, 2013. arXiv:1301.6551.

Mermin, D, (2014) "Physics: QBism puts the scientist back into science", Nature, 507, 422

Pusey MF, Barrett J, Rudolph T., (2012), "On the reality of the quantum state". Nature Physics, 8, 475